\documentstyle[prc,aps,epsfig,multicol]{revtex}
\tightenlines
\begin{document}
\title{Diomega production in relativistic heavy ion 
collisions}
\author{Subrata Pal,$^1$ C. M. Ko,$^1$ and Z. Y. Zhang$^2$}
\address{$^1$Cyclotron Institute and Physics Department,
Texas A\&M University, College Station, Texas 77843-3366}
\address{$^2$Institute of High Energy Physics, 100039 Beijing,
People's Republic of China}

\maketitle

\begin{abstract}
Using a multiphase transport model, we study the production of
a new strange dibaryon $(\Omega\Omega)_{0+}$ in dense hadronic matter 
formed in relativistic heavy ion collisions. The (multi-)strange baryons
($\Xi$ and $\Omega$) are produced by strangeness-exchange reactions between 
antikaons and hyperons in the pure hadronic phase. The rescattering between
the $\Omega$s at midrapidity leads to a production probability of 
$\simeq 3\times 10^{-7}$ $(\Omega\Omega)_{0+}$ per event 
at the RHIC energy of $\sqrt s=130A$ GeV. The production probability would
be enhanced by two orders of magnitude if $(\Omega\Omega)_{0+}$ 
and $\Omega$ reach chemical equilibrium during heavy ion collisions.
We further find that the yield of $(\Omega\Omega)_{0+}$ 
increases continuously from SPS to the highest RHIC energy.

\medskip
\noindent PACS numbers: 25.75.-q, 24.10.Lx, 14.20.Pt
\end{abstract}

\begin{multicols}{2}

Since first proposed by Jaffe \cite{jaffe} that the H dibaryon with
quark content ($uuddss$) could be a possible
lightest strangelet $-$ droplet of bound strange quark matter, 
properties of dibaryons have been widely investigated in various models,
such as the MIT bag model \cite{chodos}, the Skyrme model 
\cite{adkins,nyman}, and the constituent quark model 
\cite{oka,yazaki,valcarce,shimizu}. The consensus regarding the mass 
of H dibaryon is about 2.232 GeV \cite{dover}. However, 
extensive experimental searches \cite{imai} have not identified any 
strangelets with small values of charge fraction, $f_Z=|Z|/A <1$, 
and strangeness fraction $f_S=|S|/A < 2$. 

Recently, the structure and properties of dibaryons with large strangeness
are investigated in the chiral SU(3) model \cite{zhang00,zhangc,li01}
that have been quite successful in reproducing several nuclear properties 
\cite{zhang97}. In this model the H dibaryon is found to be only weakly 
bound \cite{li01}. On the other hand, analysis of some six-quark cluster 
states with high strangeness fraction reveals that the diomega 
$(\Omega\Omega)_{0+}$, 
in particular, is rather deeply bound. Although the color magnetic 
interaction in the one gluon exchange term for this system 
exhibits repulsive feature, the large attraction stemming from the chiral 
quark coupling and from the symmetry property of the system lead to a rather
large binding. In the Resonating Group Method calculation, the binding 
energy of $(\Omega\Omega)_{0+}$ is found to be as large as $\approx 116$ MeV,
and the root mean square distance between the two $\Omega$'s is 0.84 fm. 
Besides the large (negative) charge fraction $f_Z=1$ and strangeness fraction 
$f_S =3$, the new dibaryon $(\Omega\Omega)_{0+}$ has quite a long mean 
lifetime of $\sim 10^{-10}$ sec as it can undergo only weak decay.

Because of its large strangeness, $(\Omega\Omega)_{0+}$ is not
likely to be produced in proton-proton collisions. On the other hand, 
strangeness production is enhanced in heavy ion collisions and has
been suggested as one of the possible signals of quark-gluon plasma (QGP) 
due to large gluon density and low energy threshold for $s\bar s$ 
formation \cite{rafelski,koch}. This may thus
lead to the formation of exotic deeply bound objects composed of quarks
or baryons with large strangeness. 
Therefore, in relativistic heavy ion collisions, especially at RHIC energies, 
the dibaryon $(\Omega\Omega)_{0+}$ could be a new interesting candidate.

Indeed, recent measurements by the WA97 Collaboration \cite{sps} and the NA49 
Collaboration \cite{marg} demonstrated substantial enhancement of 
the (anti-)hyperon yields ($\Lambda$, $\Xi$, and $\Omega$) in $158A$ GeV 
Pb-Pb central collisions relative to p-Pb collisions. The enhancement pattern 
increases with the strangeness content of the (anti-)hyperon. Such a large
enhancement for multistrange baryons at midrapidity was interpreted as a 
signal for quark-gluon plasma formation.

On the other hand, even without a phase transition (anti)strangeness 
can be abundantly produced by hadron rescatterings alone. In fact,
within the microscopic transport model UrQMD, the WA97 data for multistrange
baryon enhancement can be explained by reducing the constituent quark mass
in the fragmentation of the initial strings in dense matter or by increasing
the string tension \cite{soff}. Based on the rate equation approach, the
multimesonic reactions $\bar Y + N \leftrightarrow n\pi + n_YK$ was 
demonstrated \cite{carsten} to enhance antistrange hyperon $\bar Y$ 
production in a hadronic scenario. Using a multiphase transport (AMPT) 
model, we found that strangeness-exchange reactions between antikaons 
and hyperons in a pure hadronic stage also leads to a significant production 
of multistrange baryons at the SPS and RHIC energies \cite{pal}.

The considerable production of (multi-)strange baryons in the dense hadronic
stage, in absence of QGP formation, could then also result in the formation
of exotic objects with enhanced strangeness. Properties of metastable
multistrange baryonic objects consisting of nucleons, $\Lambda$'s and $\Xi$
are studied in the relativistic mean-field theory \cite{schaf97}. These objects
were found to have properties quite similar to those of strangelets $-$
their quark counterpart. Estimates on the production of such strange 
dibaryons were presented \cite{schaf00} at the RHIC energies by 
employing wave-function coalescence in the RQMD model. 

The pronounced $\Omega$ production obtained in the AMPT model from 
strangeness-exchange reactions suggests that multiple collisions between these 
omegas in the dense hadronic matter may lead to an appreciable 
production of the new strange dibaryon $(\Omega\Omega)_{0+}$. In this letter, 
we investigate $(\Omega\Omega)_{0+}$ production in relativistic
heavy ion collisions in the AMPT model.

The AMPT model \cite{ampt} is a hybrid model that uses as input both
the minijet
partons from the hard processes and the strings from the soft processes
in the HIJING model \cite{hijing}. The dynamical evolution of partons
are modeled by the ZPC \cite{zpc} parton cascade model, while the
transition from the partonic matter to the hadronic matter is based
on the Lund string fragmentation model \cite{lund}. The final-state hadronic
scatterings are then modeled by the ART model \cite{art}.
The AMPT model has been very successful in describing the measured 
transverse momentum and rapidity distributions of charge particles 
\cite{back1,acker} as well as the particle to antiparticle ratios
\cite{back2}.
The multistrange baryon production is via the strangeness-exchange reactions
${\bar K}\Lambda\leftrightarrow\Xi\pi$, ${\bar K}\Sigma\leftrightarrow\Xi\pi$ 
and ${\bar K}\Xi\leftrightarrow\Omega\pi$. Since there is no experimental 
information on these cross sections, they are assumed to be the same as that 
for $\bar KN\to\Sigma\pi$, which is the isospin averaged cross
section for converting a nucleon to a sigma \cite{pal}.
This assumption is consistent with results obtained using 
SU(3) invariant hadronic Lagrangians \cite{li}.
For diomega production from collisions between omegas, we consider
both the the electromagnetic process 
$\Omega + \Omega \to (\Omega\Omega)_{0+} + \gamma$ and the
strong interaction process $\Omega + \Omega \to (\Omega\Omega)_{0+} + \eta$. 
Their cross sections have been evaluated in Ref. \cite{yu} using
an effective Hamiltonian.  For the process 
$\Omega + \Omega \to (\Omega\Omega)_{0+} + \gamma$, the maximum cross 
section is found to be $\approx 1.6 \mu{\rm b}$ at $p_\Omega \approx 0.5$ GeV,
while that for the process $\Omega + \Omega \to (\Omega\Omega)_{0+} + \eta$,
which requires a threshold $p_\Omega > 0.88$ GeV due to finite $\eta$ mass,
reaches a maximum value of $\sim 5.5 \mu{\rm b}$ 
at $p_\Omega \approx 1.1$ GeV.  Because of the mismatch between the large 
initial and small final relative momenta between the two omegas in the reaction
$\Omega + \Omega \to (\Omega\Omega)_{0+} + \eta$, which involves
only strong interactions, its cross section is only slightly larger than 
that for the reaction $\Omega + \Omega \to (\Omega\Omega)_{0+} + \gamma$,
which involves a much weaker electromagnetic interaction.

To include the diomega production reactions in the AMPT transport model, 
we parametrize their cross sections as functions of center of mass energy 
$\sqrt s$ (in GeV), i.e., 
\begin{equation}
\sigma_{\Omega+\Omega\to (\Omega\Omega)+\gamma} = 
87\left(1- \frac{\sqrt s_0}{\sqrt s} \right)^{\! 0.91}
\left(\frac{\sqrt s_0}{\sqrt s} \right)^{\! 24.6} ~ \mu{\rm b},
\end{equation}
where $\sqrt s_0 = 2m_\Omega =  3.345$ GeV, and
\begin{equation}
\sigma_{\Omega+\Omega\to (\Omega\Omega)+\eta} = 
1.5\times 10^4\left(1- \frac{\sqrt s_0}{\sqrt s} \right)^{\! 2}
\left(\frac{\sqrt s_0}{\sqrt s} \right)^{\! 34.9} ~ \mu{\rm b} ,
\label{diometa}
\end{equation}
where $\sqrt s_0 = m_\eta + 2m_\Omega - B_{(\Omega\Omega)} = 3.777$ GeV. 

Due to small number of multistrange baryons in heavy ion collisions, 
the productions of $\Xi$ and $\Omega$ as well as of the diomega
are treated perturbatively in the AMPT model \cite{pal}, i.e., the collision 
dynamics is assumed to be unaffected by the production of these particles.
For $(\Omega\Omega)_{0+}$ production, since its abundance
is extremely small, we assume that the production/annihilation rate of
$\eta (\gamma)$ is unaffected. Taking a total scattering
cross section between two $\Omega$s to be $\sigma_{\Omega\Omega} = 20$ mb,
the two $\Omega$s will collide if their distance of closest approach is
$\sqrt{\sigma_{\Omega\Omega}/\pi}$. If $P_\Omega$ and $P_\Omega^\prime$ are
the production probabilities of the colliding $\Omega$s, the probability of 
the resulting diomega is then 
$P_{(\Omega\Omega)_{0+}} = P_\Omega \cdot P_\Omega^\prime \cdot
\Gamma_{ \Omega+\Omega \to (\Omega\Omega)_{0+}}$, where 
$\Gamma_{\Omega + \Omega \to (\Omega\Omega)_{0+}} = 
(\sigma_{\Omega+\Omega \to (\Omega\Omega) + \gamma} +  
\sigma_{\Omega+\Omega \to (\Omega\Omega) + \eta} ) /\sigma_{ \Omega\Omega}$.
The probabilities of each of the interacting omegas are reduced by 
$P_{(\Omega\Omega)_{0+}}$.
The annihilation cross section of a diomega via the reaction 
$(\Omega\Omega)_{0+} + \eta \to \Omega + \Omega$ is obtained using detailed
balance of Eq. (\ref{diometa}). If a diomega collides with a eta, the
diomega is totally annihilated and two omegas are produced with isotropic
momentum distribution and each having a probability corresponding to the 
diomega production probability of $P_{(\Omega\Omega)_{0+}}$.
Since diomega production is treated perturbatively, the final results 
do not depend on the value of $\sigma_{\Omega\Omega}$ used in the
AMPT model.

\begin{figure}[h]
\centerline{\epsfig{file=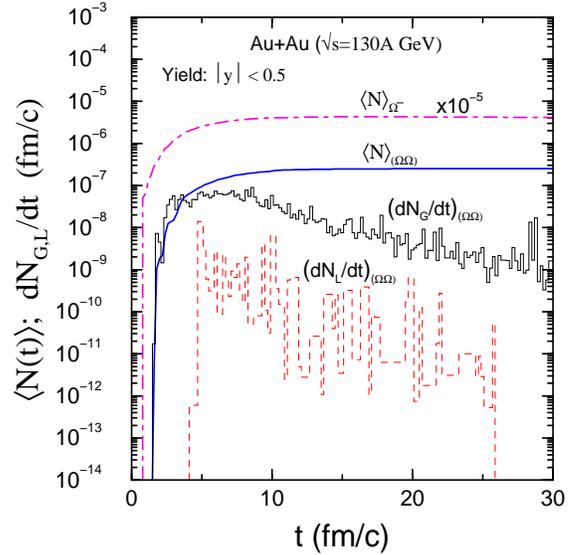,width=2.9in,height=2.9in,angle=0}}
\vspace{0.12cm}
\caption{Time evolution of midrapidity $\Omega^-$ and $(\Omega\Omega)_{0+}$ 
for Au+Au collisions at the RHIC energy of $\sqrt s = 130A$ GeV at an impact 
parameter of $b\leq 3$ fm in the AMPT model. The production and absorption
rates for the dibaryon $(\Omega\Omega)_{0+}$ as functions of time is also
shown.}
\label{time}
\end{figure}

In Fig. \ref{time}, we show the time evolution of the abundances of 
midrapidity 
$\Omega^-$ and $(\Omega\Omega)_{0+}$ in central Au+Au collisions at the 
RHIC energy of $\sqrt s=130A$ GeV. As is evident from the figure, most 
of the omegas and $(\Omega\Omega)_{0+}$ are produced within 10 fm/c 
after hadronization when the energy density is high. Beyond this time 
the diomega yield saturates to a value of $2.8 \times 10^{-7}$
per event. 

From the study of time evolution of the production and absorption rates of 
particles, we found \cite{pal} that most strange hadrons, including the 
$\Omega$s, reach chemical equilibrium in collisions at the RHIC energy 
of $\sqrt s =130A$ GeV. In Fig. \ref{time} we depict the time evolution of the 
production (solid lines) and absorption rates (dashed lines) of
$(\Omega\Omega)_{0+}$. It is observed that the absorption rate of a diomega
by colliding with a eta meson is always smaller compared to its production 
rate at all times. This stems from additional cross section in the production 
channel due to the electromagnetic process which has no threshold. We thus 
infer that the small number of diomega $(\Omega\Omega)_{0+}$ that may be
produced in the relativistic collisions at RHIC do not reach chemical 
equilibrium.

\begin{figure}[h]
\centerline{\epsfig{file=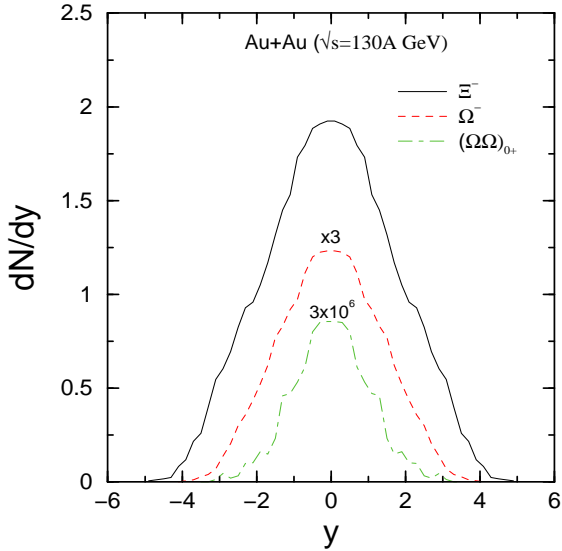,width=2.9in,height=2.9in,angle=0}}
\vspace{0.12cm}
\caption{Rapidity distributions of $\Xi^-$, $\Omega^-$, and 
$(\Omega\Omega)_{0+}$ for Au+Au collisions at the RHIC energy of 
$\sqrt s = 130A$ GeV at an impact parameter of $b\leq 3$ fm in the AMPT model.}
\label{rapidity}
\end{figure}
\vspace{0.2cm}

\begin{figure}[h]
\centerline{\epsfig{file=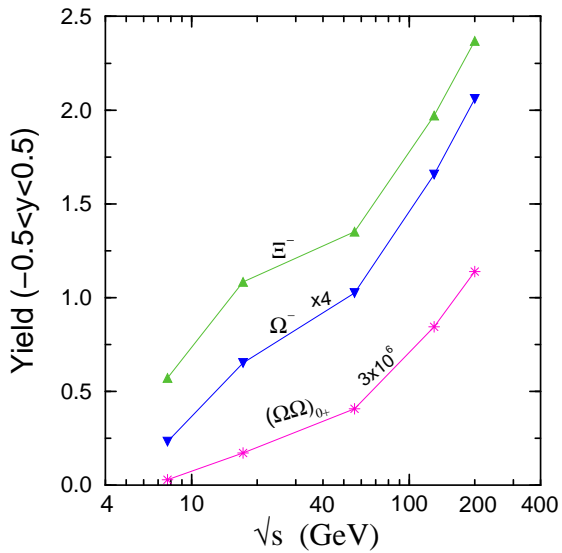,width=2.9in,height=2.9in,angle=0}}
\vspace{0.25cm}
\caption{Energy dependence of $\Xi^-$, $\Omega^-$, and $(\Omega\Omega)_{0+}$
at midrapidity $|y|<0.5$ for heavy ion collisions at an impact parameter
of $b\leq 3$ fm in the AMPT model.} 
\label{energy}
\end{figure}

The rapidity distributions of the produced hadrons peak at midrapidity 
\cite{ampt} because of increased multiple collisions and rescattering 
at central rapidities compared to that at large rapidities.  In Fig.  
\ref{rapidity} we
show the rapidity distribution of the multistrange particles $\Xi$ and 
$\Omega$, along with that of $(\Omega\Omega)_{0+}$. The 
rapidity distribution of particles with increasing strangeness content 
gradually become narrower as these particles are successively produced from
their parent that collide more frequently at small rapidities where the baryon 
density is high.

We have also studied the energy dependence of diomega production in 
relativistic heavy ion collisions. In Fig. \ref{energy}, we show the
yield of $(\Omega\Omega)_{0+}$ at midrapidity in Au+Au collisions
at various energies. Also shown in the figure are the yields of
$\Omega^-$ and $\Xi^-$ as functions of energies. It is seen
that while the $\Omega^-$ number increases by about a factor
of three from the SPS energy ($\sqrt s =17A$ GeV) to the highest
RHIC energy ($\sqrt s =200A$ GeV), the $(\Omega\Omega)_{0+}$ 
number increases by about a factor of 4, indicating that the
dibaryon $(\Omega\Omega)_{0+}$ number reveals a much faster rate
of increase with $\sqrt s$ as compared to $\Omega^-$. 

The AMPT model gives a lower bound on the diomega production probability
since there might be other production channels, such as 
$\Omega^- + \Xi^- \to (\Omega\Omega)_{0+} + K^0$, which have been neglected 
in the present calculation because of their unknown cross sections.
If the diomega production cross section is larger, then diomegas 
may reach chemical equilibrium with omegas in heavy ion collisions.
For most other particles including multistrange baryons, equilibrium
thermal models have been successfully employed \cite{braun,cley} 
to explain experimental data for their yields and ratios over a wide 
range of energies from AGS to RHIC. Adopting 
the statistical model that is based on the grand canonical ensemble
with complete thermal, chemical, and strangeness equilibrium,
we have found that the results from the AMPT model for the $K^+/\pi^+$,
${\bar p}/p$, and $K^-/K^+$ ratios, which are about 0.18, 0.65, and
0.89, respectively, at midrapidity in Au+Au collisions at
energy $\sqrt s = 130A$ GeV, can be approximately 
described with a temperature $T\simeq 170$ MeV, baryon chemical 
potential $\mu_B\simeq 37$ MeV, and strange
chemical potential $\mu_S\simeq 10$ MeV.  If we assume that diomegas are 
also in chemical equilibrium with omegas, then 
the ratio $(\Omega\Omega)_{0+}/\Omega^-$ is $7.4\times 10^{-5}$.
With the omega number of about 0.41 at midrapidity, this leads to a diomega
production probability of $\sim 3.0\times 10^{-5}$ per event, which is
two orders of magnitude higher than that obtained in 
our transport model. 

Diomega production in heavy ion collisions can also be studied 
using the coalescence model \cite{schaf00,sato,gyul} based on the
omega phase space distribution at freeze out as obtained in the AMPT model.
Assuming a harmonic oscillator wave function \cite{gyul}, the Wigner 
density for the diomega is $\rho_D({\bf r},{\bf q}) = 8 \: 
\exp(-r^2/d^2 - q^2d^2)$, where
${\bf r}={\bf r}_1 - {\bf r}_2$ and ${\bf q}=({\bf p}_1 - {\bf p}_2)/2$
are given in the c.m. system of $\Omega$-$\Omega$, and the value of $d=0.69$ fm
corresponds to a rms radius of 0.84 fm for the diomega. For central
Au+Au collisions at $\sqrt s = 130A$ GeV, the yield of $(\Omega\Omega)_{0+}$ 
at midrapidity is found to be $2.6\times 10^{-5}$ and is comparable 
to that from the thermal model.
The similar results obtained from both the thermal and coalescence models are
not surprising as it was shown in Ref. \cite{mekjian}, 
that the two models are equivalent when 
matter is in thermal and chemical equilibrium and the binding 
energy of the composite particle is much smaller than the temperature. 
Since the AMPT model predicts a hadronic matter 
at freeze out that is close to thermal and chemical equilibrium, 
the diomega yield from the coalescence model thus should be similar to 
that given by the thermal model.

Our estimate for the production probability of $(\Omega\Omega)_{0+}$ 
are well within the limits of the present detectors 
used at RHIC energies. Therefore, this exotic object can, in principle, be
detected in present and future experiments.
The fact that $(\Omega\Omega)_{0+}$ has a large strangeness content with a
binding energy of $\simeq 116$ MeV, it is stable against strong hadronic
decays and possess weak decays:
$(\Omega\Omega)_{0+} \to \pi^- + \Xi^0 + \Omega^-$ and 
$(\Omega\Omega)_{0+} \to \pi^0 + \Xi^- + \Omega^-$. Because three-body 
decay are involved, the final state phase-space would be suppressed.
In the sudden approximation, the mean lifetime of
$(\Omega\Omega)_{0+}$ was found \cite{zhangc} to be about four times 
longer than the free $\Omega$ lifetime of $0.822 \times 10^{-10}$ sec. 
Apart from these conventional decay modes, the nonmesonic decay 
$(\Omega\Omega)_{0+} \to \Xi^- +\Omega^-$ is also possible; 
the estimated \cite{zhangc} lifetime of $(\Omega\Omega)_{0+}$ 
for this process is twice the free $\Omega$ lifetime. Thus, instead of
direct observation, the $(\Omega\Omega)_{0+}$ may also be detected 
in the $\Xi^-\Omega^-$ invariant mass distribution. The observation of which
could provide useful information of the unknown $\Omega-\Omega$ 
interaction strength. Our study thus opens up the intriguing possibility 
of detecting the new dibaryon $(\Omega\Omega)_{0+}$ in
heavy ion collisions at the RHIC energies.
\vspace{0.2cm}

This work of SP and CMK is supported by the National 
Science Foundation under Grant No. PHY-9870038, the Welch Foundation 
under Grant No. A-1358, and the Texas Advanced Research Program under 
Grant No. FY99-010366-0081, while that of ZYZ is supported by
the National Natural Science Foundation of China and the Chinese 
Academy of Sciences.

\end{multicols}

\end{document}